# Chemicals in the Creek: designing a situated data physicalization of open government data with the community

Laura J. Perovich, Sara Ann Wylie, Roseann Bongiovanni

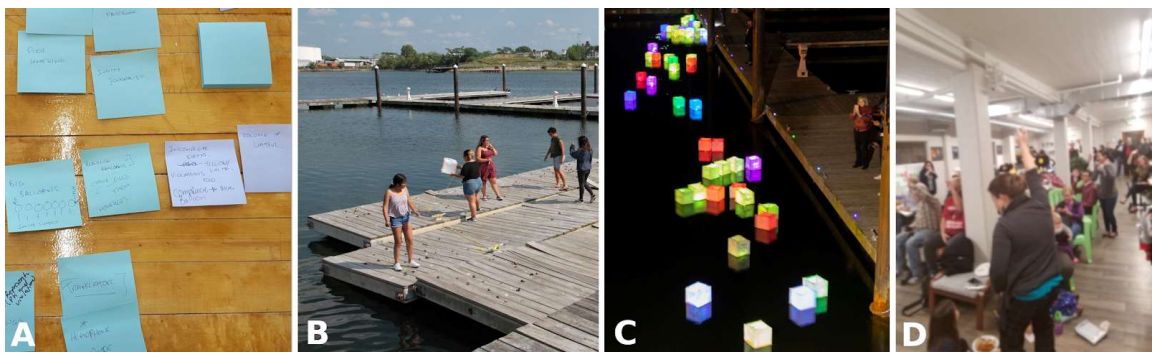

Fig. 1.Chemicals in the Creek is a situated data physicalization that aims to make open governmental data accessible and meaningful for non-expert users. The event was (a, b) designed and created collaboratively with community partners in Chelsea, Massachusetts, (img: Sara Wylie, Garance Malivel) (c) held at night on the Chelsea Creek (img: Will Campbell), and (d) focused on local action (img: Sara Wylie).

**Abstract**—Over the last decade growing amounts of government data have been made available in an attempt to increase transparency and civic participation, but it is unclear if this data serves non-expert communities due to gaps in access and the technical knowledge needed to interpret this "open" data. We conducted a two-year design study focused on the creation of a community-based data display using the United States Environmental Protection Agency data on water permit violations by oil storage facilities on the Chelsea Creek in Massachusetts to explore whether situated data physicalization and Participatory Action Research could support meaningful engagement with open data. We selected this data as it is of interest to local groups and available online, yet remains largely invisible and inaccessible to the Chelsea community. The resulting installation, Chemicals in the Creek, responds to the call for community-engaged visualization processes and provides an application of situated methods of data representation. It proposes event-centered and power-aware modes of engagement using contextual and embodied data representations. The design of Chemicals in the Creek is grounded in interactive workshops and we analyze it through event observation, interviews, and community outcomes. We reflect on the role of community engaged research in the Information Visualization community relative to recent conversations on new approaches to design studies and evaluation.

**Index Terms**—data physicalization, Participatory Action Research, water quality, environmental HCI

✦

## 1 INTRODUCTION

Over the last 15 years there has been a large shift towards open government data often provided through online tools, such as the European Union's Open Data Portal [67]. In many cases, this data sharing is described as a way to "establish a system of transparency, public participation, and collaboration" in governments through providing more information to the public and has become an official part of policy or law [65, 66]. These resources have proved valuable to researchers [28], industries [81], and journalists [37] and some non-profit organizations [7]. Yet this data may remain out of reach for those who do not have substantial technical knowledge, subject area expertise, or awareness of the data, even when it is highly relevant [14, 26].

- *Laura J. Perovich is with Art+Design at Northeastern University and the MIT Media Lab. E-mail: perovich@media.mit.edu.*
- *Sara Ann Wylie is with Sociology/Anthropology and Health Sciences at Northeastern University. E-mail: s.wylie@northeastern.edu.*
- *Roseann Bongiovanni is with GreenRoots, Inc. E-mail: RoseannB@greenrootschelsea.org.*

This study uses a performative situated data physicalization created using Participatory Action Research (PAR) to engage communities with open government data. It aims to facilitate community action through increasing the accessibility and visibility of the data using a collective experience. We use data from the United States (US) Environmental Protection Agency (EPA) Enforcement Compliance History Online (ECHO) database [78] which compiles industry self-reported monitoring information based on National Pollution Discharge Permits (NPDES) which allow industrial facilities to discharge permitted quantities of waste into bodies of water as part of the Clean Water Act (CWA). We worked on the Chelsea Creek which is zoned for industrial use by Massachusetts and is home to seven of the region's oil storage facilities which hold much of the regional heating oil and all of the jet fuel for Boston's airport [49]. Neighboring the river is Chelsea, MA, a small city that meets the EPA's environmental justice criteria [79] and whose majority low-income, non-English speaking residents bear a disproportionate burden of industrial risks and environmentally related illness like asthma [31]. Our PAR collaboration with GreenRoots—an environmental justice organization in Chelsea—and their youth group the Environmental Chelsea Organizers (ECO), led us to develop a public performance of local NPDES water permit violations by oil storage facilities from 2013-2017, described here.

**Participatory Action Research (PAR):** PAR has been used for over 80 years and has an extensive theoretical and practical history

[9]. "Defined most simply, PAR involves researchers and participants working together to examine a problematic situation or action to change it for the better" [40]. A key purpose of PAR is to engage scientists and communities "in making sense of real-life situations and acting on them" through a research process [9]. PAR does research "with and for, rather than on, participants" [40]. Briefly, the three main pillars of PAR are:

- **Participation**: PAR is a "socially owned process" that uses democratic means to examine "life in society" [9, 40]. Building of trust between researcher-community teams is central. It rejects a deficit model of learning in which a teacher informs a student about their conditions and instead it aims to build co-learning capacity on the team [9, 23].
- **Action**: PAR is "concerned first and foremost with supporting social change in complex settings" [9]. It focuses on experience and impact on real-world challenges particular to a local community context [40, 51]. It frequently engages with issues of power in the context of structural oppression [51].
- **Research**: PAR integrates "the advancement of knowledge" with participation and action [9] and often includes very interdisciplinary approaches. PAR gives significant value to the process of research and projects may be seen as "change experiment(s)" [9].

In practice, PAR includes "a variety of participatory approaches to action-oriented research" [40] including connections to Design Thinking, Sociotechnical Systems Theory, Experiential Learning, and Critical Pedagogy, among many others [9]. It also has addressed a wide range of issues such as "education, the workplace, community life, intermediate technologies, [and] environmental degradation" [9]. On the ground, PAR often uses hands-on methods such as group work, community art, political action, learning by doing, and storytelling [40]. Methods are selected with the community to respond to local needs and preferences.

PAR approaches fit within recent calls for a "new, interpretivist perspective on visualization design study that extends and deepens the existing definition" [52]. Meyer et. al.'s perspective is informed by work in social science—some of which is the backdrop of PAR methods—and includes discussion of action research, a relative of PAR [52]. Like PAR, this design research approach "is inherently messy, changing, subjective, and context specific" and involves considerable efforts to "build trust, develop agency, and invite interest in the design process" [52]. PAR expands these ideas slightly by centering the importance of creating action on social issues in the community's context, not only the research setting. PAR also prompts us to further probe the idea of partnership with a "domain expert" in design studies. While communities may be considered "domain experts," the traditional interpretation of this role in design studies may limit our imagination of what this partnership looks like and not do justice to the engagements typical in PAR.

Ongoing methods in data visualization and HCI, such as co-design and in-the-wild approaches, have some overlap with PAR and a few projects in these areas may fall under the PAR umbrella. Yet, these methods do not inherently include all the priorities and values of PAR. In-the-wild methods "refer to research that seeks to understand and shape new technology interventions within everyday living" and are generally technology-first though they do consider the context and social implications of technologies [13]. Co-design centers the process of making through broad participation and "refers to any act of joint creativity in which designers, end-users untrained in design, and perhaps other stakeholders work together in design processes" [60], [83]. PAR goes beyond these approaches by centering local social action, co-learning, and issues of power.

**Project summary, aims, and contributions:** Chemicals in the Creek builds on work in community based data physicalization and draws on PAR approaches to respond to the call for civic data display processes that are engaged with communities throughout the research process and deal with challenging social issues [12]. Our PAR research aim was not only to better understand that data but also to create an action that changed the balance of power in productive ways for GreenRoots relative to the oil storage facilities. We use a performative, collectively experienced, situated, and embodied approach oriented towards action. This stands in contrast to more common approaches in the field where data displays are made available over a significant period of time or are designed for individual interaction, asynchronous interaction, chance encounters, or remote interaction. This work also contributes to expanding the methodology landscape for Information Visualization research by building from new project evaluation and design research approaches [52, 70, 86] and proposing a role for PAR methods.

## 2 RELATED WORK

**Situated data displays:** Data physicalization research has begun to play an important role in the Information Visualization community [35, 36, 73] and has provided opportunities for contextualizing information in the environment, building on tangible interfaces [87], and contributing to participatory fabrication and visualization pedagogy [4, 32, 55, 76, 77]. Central to our work is the model by Willett et al., that describes the data display space along two axes: the medium ranging from visualization to physicalization, and the level of integration of the data into the context it came from, from not integrated to situated to embedded [87]. Along the media axis, data physicalizations are "a physical artifact whose geometry or material properties encode data" [36], though the transition from visualization to physicalization is likely more continuous and includes displays that are physically manifest in the world but are primarily engaged with through limited material properties [87]. Researchers have outlined a number of possible benefits of data physicalization, including in "making data accessible," "bringing data into the real world," "engaging people" [36], "supporting collocated collaboration" [87], and "encouraging reflection" [76, 77]. Situated methods of representing data may "allow viewers to examine and extract additional information not present in the dataset itself" [87] and leverage a physical environment's "potential to act as an information-carrying medium" [12].

**Community-based participatory data displays:** Researchers have explored various types and amounts of community participation in creating and interacting with public data displays [30]. Visualising Mill Road uses urban infrastructure, such as sidewalks, as a canvas for low-tech data displays that provide local information and employ participatory methods for data collection [42]. Public visualizations and polling systems allow community members to explore and contribute to data displays that speak to local issues [10, 29, 74]. Claes et al. compare tangible, digital, and mixed ways of interacting with civic data through displays in the urban environment and find that physical approaches to sharing data can lead to more interaction and deeper insights [11]. They also underline the importance of attention to "meaningfulness, relevance and timeliness within the social, spatial and cultural realities of its immediate environment" when selecting data for public display [11], pointing towards one benefit of community processes that emphasize social engagement.

Schoffelen et al. reflect on ways visualizations can engage diverse communities with complex civic issues and proposes three core aspects of readability in civic visualizations—engagement, sense making, and reflection [69]. They underline the importance of incorporating ways for people to contribute their perspectives and suggest that long-term participatory design may be used to gather people for conversation and disagreement [69]. Street Infographics uses street sign displays and graffiti to share information with a general population and to contextualize data displays and blend them seamlessly with existing aesthetics [12]. Importantly, Street Infographics researchers observe that urban visualizations can have unexpected community impacts and may be mistrusted or misinterpreted if they surface sensitive issues [12], causing the researchers to recommend community design processes. Chemicals in the Creek responds to this invitation by employing a PAR process that deeply involves the community throughout the research study and centers their local expertise. PAR is well positioned to contend with questions of power tied to difficult social issues like immigration that surface in urban data display research [12, 69] and are less embedded in many participatory design practices.

**Public displays of environmental information:** The integration of our work in the cityscape draws on research on urban data displays. Tidy Street uses individual homes to publicly display energy use to promote conversations within neighborhoods [5]. Gough et al. describe InterANTARCTICA and Reefs on the Edge, two interactive museum-based visualizations of environmental challenges that focus on the importance of affect in engaging non-expert users and moving towards change on these important issues [25]. A Conversation Between Trees similarly focuses on affective modes of engagement around environmental data using art, as well as physical and situated modes of display [34]. Particle Falls created an artistic interpretation of real-time air quality data and projects it onto a public building in order to raise awareness about air pollution and its health impacts [63]. Climate Prisms: The Arctic takes an artistic approach to engaging people with environmental issues in hopes of deepening understanding and motivating change through engaging with difficult topics [68]. These public visualizations may be designed "to promote awareness, discussion and participation" within communities [82] and can be used to support easier community access to the information [25] and build human connection to the environment [68]. Chemicals in the Creek uses the city canvas for sharing information and takes a more active approach to facilitating community engagement through creating a performance instead of an ambient information display. This builds on work in HCI on performance as research, particularly in its potential to build trust and connection on difficult topics [75].

**Environmental data physicalization:** A number of environmental data physicalizations have emerged from HCI, information design, and arts communities. WearAir provides a semi-public interface for air quality information using a LED display built into a t-shirt [38]. Air Quality Balloons provide a more public display of air quality information by using interactive color changing balloons to reflect real-time air quality data. It is an example of spectacle computing, which "vibrantly project[s] information into the public sphere using expressive and tangible media" [43]. Feral Robotic Dogs invites communities to hack toy robotic dogs so that they "sniff" out air pollution and swarm towards polluted areas in public spaces [45]. In autographic data representations, the act of capturing the data is the display itself [56]; in the Garden of Eden lettuce is grown in sealed boxes filled with air matching the pollution level in different cities to show us the impact of that air pollution [57]. Other environmental citizen science research focuses on systems for community engagement and data gathering that may or may not integrate visualizations of the results [1, 19]. Chemicals in the Creek draws on spectacle approaches to environmental data but uses publicly available government data instead of real-time sensor data. This difference is significant since governments may be more responsive to community requests founded in government data, as many agencies do not currently have mechanisms for including citizen science data in their assessments [39, 58].

**Visualization tools for open data and the public:** There are some visualization tools available for non-expert users. Governments may provide basic visualization tools within their open data portals, for example, the US EPA's Outdoor Air Quality site [80]. Researchers have created tools to help investigative journalists mine and visualize Freedom of Information Act documents [8] and understand how non-experts interact with open data visualization tools [6]. More general data visualization tools, such as Many Eyes, may be accessible to individuals with varied experience and contribute to the democratization of visualization [83]. Constructive visualization uses physical approaches to create visualizations that are simple, expressive, and dynamic [32] and approaches to community-based open data visualization include programs for improving data literacy in the context of Smart Cities [88] and in school programs [46].

## 3 METHODS

### 3.1 Research Methods

Our research methods are informed by recent work by Wang et al. that proposes an expanded assessment of the value of visualizations that takes into account that "emotional qualities of experiences are part of sense-making" [50, 86]. This "holistic approach" centers the need to consider hedonic qualities and values in a way that goes beyond the "[traditional] focus on efficiency, comprehension, or insight" [86]. Specifically, the value of a data display includes user engagement through four means:

- **Affective**: emotional engagement, such as "feelings of awe,...wonder...amusement, concern...[or] anger" [86],
- **Physical**: engagement through "touch or movement; real or imagined" [86],
- **Intellectual**: includes "activities such as recognition, analysis, and contemplation." This may include, but is not limited to, traditional efficiency measures in visualization [86],
- **Social**: engagement between people, including conversation as well as embodied engagement such as "laughing, gesturing, and mimicking the body postures of others" [86].

These considerations dovetail with work in data physicalization and data art where many projects aim for emotional engagement, and research in cognitive science that "tells us that emotion and physical touch can be tied directly to motivations, actions, and learning" [86]. They also align with Meyer et al.'s "reposition[ing] [of] design study as a rich, subjective, and interpretive approach to visualization research inquiry" by including a larger understanding of what information is important in understanding the impact of a data display [52]. This broader approach "could enable the community to address new uses for visualization and open itself to new practitioners and researchers from diverse backgrounds" [86].

Our data collection practices (detailed in 3.2) focused on capturing these parameters using research methods that follow the values of PAR and also on surfacing "action" outcomes important to this research method. To do our PAR research "with and for, rather than on, participants" [40] we were careful to use lightweight approaches to data collection that would not disrupt the action and participation aims of the project. We also considered both the final event and the research process. These approaches are further supported by Meyer et al.'s recommendation that design study researchers should represent their process and take caution to select methods of recording data that are "not disruptive" [52]. To respect the time constraints of community members, we integrated data collection into the community event when possible, as we recognized that follow-up data gathering could be burdensome.

### 3.2 Data

We collected data throughout the research process and the Chemicals in the Creek event. Descriptions of the groups, including group identifiers used for interview excerpts, are detailed in Table 1. Ethnographic and interview data include: photographs, videos, researcher reflections, observations (design process and event); technology artefacts, group reflections, over 25 workshops plans and outcomes over two years (design process); and event planning documents, community post-it note responses to the event, newspaper articles, community reactions, semi-structured interviews (event). We have included over 75 photos and videos and additional documents in the Supplemental Information (figures denoted with the letter S) to provide supporting details, as suggested in Meyer et. al. [52].

Post-it note feedback during the event asked what was good about the data physicalization, how it could be improved, what questions the event raised, and the future of the Chelsea Creek. Semi-structured interviews after the event ran from 20 minutes to over an hour and explored experiences of the event, strengths and weaknesses of the physicalization, emotional resonance and social engagement, and the personal and collective impact.

Most event observers were on the GreenRoots mailing lists, had participated in past GreenRoots community meetings, and/or were Chelsea residents. We limited our audience to this group to meet the overall aims of GreenRoots to engage their membership with these issues and to ensure that everyone could actively participate in the conversation. At the same time, this approach may introduce bias, as this group may begin from a positive baseline affect, may have existing social connections to others at the event, and/or may have some prior knowledge of local challenges. We were able to address this some by including interviews with event volunteers who did not live in Chelsea or have knowledge of the area. We also framed interview questions to encourage critical assessments—for example, reassuring people that constructive negative feedback would help improve our future work.

### 3.3 Ideation

Our collaboration began in 2017 with four months of weekly workshops with GreenRoots' ECO youth group. These workshops included many short, collaborative, hands on, interdisciplinary projects (Figs. S2.01-07). The phase of work aimed to facilitate collective learning and trust-building on the team that could lead to the development of possible collaborative projects on water quality and environmental justice issues in Chelsea, MA. At the end of these workshops, we gained funding to implement one of these collectively developed ideas in the following year; specifically to create a display of the water violation data from the oil storage facilities on Chelsea Creek. Summarizing the local NPDES data was led by the researchers and discussed with ECO and GreenRoots. Shortly after, 5 of the 6 ECO high school student class graduated, leading us to build relationships with the new ECO group in 2018.

### 3.4 Implementation

Actualizing the installation involved decision making around the properties of the data display and creating the physical tools necessary to actualize it. The first steps included workshops with ECO and GreenRoots to understand the local importance of the oil storage facility violations, unpack the particular violations we found in the data, and surface the main messages GreenRoots wanted to convey to the community. Next, the researchers suggested a few possible ways to encode the data that reflected these priorities and selected a basic framework with feedback from ECO.

We then worked to develop physical objects that actualized that data encoding. The start of this work was conducted by the researchers and the design team, and included testing floating objects and collecting potential materials for lantern fabrication. Videos and images from this process were regularly shared with ECO in workshops to gather their feedback which informed the decision of what objects to use to represent water quality violations in the installation (Figs. S2.08-15). The second phase of lantern construction was led by the researchers and a member of the design team who was a high school student living in Chelsea and regularly worked with ECO members. This process involved extensive iteration on the materials for the lantern, the lantern design, and the lantern fabrication process, as well as the development of ways to light the lanterns. ECO students regularly participated in and gave input on the tests of the lanterns on the Chelsea Creek through workshops (Fig. 1b; Figs. S2.29-31), including decisions on mapping LED colors to the chemical variable values (Fig. S2.28). Researchers and members of the design team finalized the lantern fabrication process.

The final phase of developing the physical objects for the events was done by the researchers, ECO, and the design team. This included designing an event brochure, and building the lanterns, the project brochures, and the LED light boards at scale. Materials for the lanterns that involved specialized tools (for example laser cutters) were prepped by the researchers and then the researchers, ECO, and the design team built them together over several workshops (Fig. S2.36-39). The LED light boards were built by researchers and members of the design team and the event brochures were built by ECO, researchers, and the design team (Fig. S2.43).

The implementation phase also included developing the overall event experience. This process was led by ECO and GreenRoots in conversation with the researchers. It included selecting a time and location for the event, advertising the event, and outlining the flow of the performance in the space. ECO's knowledge of the intended audience—Chelsea residents and GreenRoots members—and the local spaces were crucial for this phase of the work.

Table 1: Background information on participating groups. Written replies (C#) are labeled per post-it note.

| group | quote ID | # people | data | age groups | description & relation |
|---|---|---|---|---|---|
| GreenRoots & ECO | E# | 16 | 6 semi-structured interview participants; ethnographic data from workshops | high school students, adults | GreenRoots is an environmental justice non-profit in Chelsea. ECO is a youth group within GreenRoots. |
| researchers | -- | 2 | ethnographic data from workshops | adult, graduate student | Researchers at local universities. Do not live in Chelsea. |
| design contributors | D# | 8 | 5 semi-structured interview participants; ethnographic data from event & process | students (undergraduate, graduate, high school) | Local students who contributed to the design process but are not part of GreenRoots. Only one lives in Chelsea. |
| event volunteers | V# | 21 | 9 semi-structured interview participants; ethnographic data from event | students (undergraduate, graduate), adults | Students or affiliates of local universities. Many have an interest in environmental issues but were not deeply involved in this project and do not live in Chelsea. |
| event observers | C# | 60 | 44 written replies; ethnographic data from event | children, students, adults | Mostly Chelsea residents. Many have attended previous events or monthly community meetings led by GreenRoots. |

## 3.5 Deployment

Chemicals in the Creek (Fig. 1c) occurred the evening of November 8th, 2018 on a dock on the Chelsea Creek as part of GreenRoots' regularly scheduled community meetings. The installation was co-led by ECO and the researchers, with contributions from the design team, and included organizing materials, coordinating event volunteers, and executing the event. The event volunteers helped put the lanterns in the water during the performance, distribute the brochures, and take pictures. The event was introduced by an ECO member and a researcher, and the community discussion after the installation was led by GreenRoots with researcher contributions.

## 4 FINDINGS: DESIGN

### 4.1 Ideation

The first phase focused on collective learning, trust building, and project development. This "fuzzy front end," detailed in [59], where the problem space was only broadly defined was crucial to ensuring that our team had the opportunity to develop project aims together; the "participation" branch of PAR. Early workshops collaboratively tested citizen science tools which helped the team discuss ways to communicate environmental data (Fig. S2.01). We developed a number of new collaborative activities during the brainstorming process (Fig. 1a; Fig. S2.02, S2.06-07) that were led by ECO; for example, we created a video pitch to add an environmentally themed PokeStop in a nearby park (Fig. S2.05).

During these workshops, our community partners reported that they were aware of EPA water monitoring data from local industries that was relevant to their community efforts, but that they did not use the data as it was difficult to locate, interact with, and interpret. The researchers produced a preliminary analysis of this data from the EPA's ECHO database [78] using traditional analysis tools (the statistical software R) and found that local oil storage companies were regularly in violation of their NPDES permits, including some violations far above the permit limits [60]. This led our team to return to an idea from earlier brainstorming and we decided to create a public data display of the violation data as part of a community event where GreenRoots could get feedback on how to integrate this information into their community work [59].

Our next step was to gain a deeper understanding of the NPDES violations from the oil storage companies on the Chelsea Creek from 2013-2017. GreenRoots' relationships with local government agencies and industries helped us collaboratively find answers to data questions. We also explored the usability of ECHO in a graduate student class "Community Based Participatory Research" at Northeastern University where we asked students to collect violation data for the oil storage facilities and summarize them by facility, year, and chemical. The students were able to complete the task with some guidance but the errors in their results indicated that the data was unlikely to be easily accessible to a general audience in its current form. Common errors included mistaking the identity of the facility and miscounting violations by using tables that indicated whether or not violations had occurred in that quarter, but did not indicate how many violations occurred.

### 4.2 Implementation

**Data Physicalization Design:** Next, we began to develop Chemicals in the Creek based on these new understandings of the data and the priorities of community partners. Our design work proceeded in four intertwined threads: design metaphor, task & data abstraction, form factors, and experience design.

*Design metaphor:* The researchers considered a number of performative metaphors, based on discussions with ECO about the overall project aim to create a legible experience around open data that could provide a space for connection and reflection, as a step towards collective action. "Ceremony" best embodies the desired attributes of the event as it creates a reflective space for observing and participating, is tied to transformation, centers connection, and builds from theoretical work on ceremony as creating shared spaces for recollection and expression of care [44].

*Task abstraction & data abstraction:* Our task abstraction formalized the project idea and values developed with ECO in "4.1 Ideation": (1) community members would be motivated to learn more about pollution in their local waterways, its impact on the community, and effecting change, (2) community groups would be motivated to engage in these conversations with local residents and use events to put pressure on polluters, without destroying the working relationships that are crucial to their efficacy in the community. To achieve this, we aimed to create an experience that was memorable, easy to engage with, and integrated into collective discussions. Artistic and physical approaches could help us capture attention and engage with a diverse audience [25, 68] and prompt social interaction in urban spaces [12]. Situated and embodied methods further connected people with the physical implications of pollution and the environments and people it impacts [15, 69, 87].

Our data abstractions were based on the relevance of the information in the community context, established through the ideation process. GreenRoots was most interested in communicating how many violations the oil storage facilities were cumulatively putting into the Chelsea Creek. This built on their ongoing advocacy efforts to get the EPA to consider the overall, instead of individual, impact of these facilities on the river and the community [79]. The relative importance of the attributes of the data were: (1) the number of violations, (2) the types of chemicals in violation, (3) the facilities the violations came from, (4) the timeline of the violations, (5) the amount of the violation relative to the permit limits. From this prioritization, we decided that each violation would be represented by an individual object, the colors of the objects would encode the chemical of that violation, and that each object would be labeled with the facility name and objects from the same facility would be grouped together. The objects would be presented sequentially to represent the time and the detailed information would be visible when holding the object. This structure is inspired by the original visual information seeking structure in digital data visualization (overview first, zoom and filter, details on demand) (Fig. S2.52) [72]. The amount of the violation was of low importance—and therefore was not physically encoded through object size—as we wanted to emphasize that each violation was in breach of a permit and therefore of significance. Additionally, comparing violation amounts may be misleading, as each chemical has different environmental and health implications, risks of chronic and acute exposures vary across chemicals in complex ways, and permit limits are established based on both the availability of technologies to avoid the pollution and on the environmental harms it can cause.

*Form factors:* We experimented with many floating form factors for representing the violations, paying attention to both functionality and appearance. As the installation was environmentally themed, we began by exploring recyclable or biodegradable materials including compostable plates, water soluble paper, candles, and more eco-friendly plastics. Unfortunately, our testing concluded that they were expensive, not robust, and hard to source so we moved our focus to creating artefacts that were reusable. Ultimately, our testing included over 15 commercial products such as floating LED candles, balloons, and paper lanterns with video documentation, detailed in the Supplemental Information. We eliminated a number of objects because they would not be visible, they would not endure "real world" river conditions such as wind, or they were too expensive.

This testing surfaced the tension between the negative nature of water pollution violations which might suggest unattractive metaphors like trash and the positive norms of ceremony that suggest more pleasant characteristics like carefully curated. While it was important not to imply that water quality permit violations are positive, we elected to use objects more aligned with the ceremony metaphor to avoid giving the impression that our team was adding trash to the waterway. We decided to hold the data physicalization at night so we could leverage existing associations between lights, ceremony, and reflective moments such as candle light vigils or Chinese lantern festivals. These considerations led us to represent each violation as a square floating lantern with an internal LED for color control, made in house by combining attributes of a number of tested commercial products (Fig. S2.52) and documented through fabrication tutorials (Figs. S1.24-27) [60]. The final lanterns were assembled with ECO at GreenRoots as another path to an embodied understanding of the data [77]. We also created an event brochure that acted both as a situated visualization during the event and a non-situated visualization after the event that would be an enduring reminder of the experience (Figs. S1.09-10, S1.29; [61]).

Our process for assigning colors to water quality violation chemicals surfaced challenges in translating between visualization best practices and the constraints of the physical world encountered in physicalization research. To reduce costs and ensure that colors that could be changed easily in the field, we designed low-cost LED color boards for the lanterns that used mechanical switches to individually control the red, green, and blue color channels on a set of tricolor LEDs, giving us seven colors (red, green, blue, light blue, yellow, pink, white). We prioritized the color assignments of the chemicals most represented in the data (TSS and pH), as those colors would contribute most to participants' overall experience of the data, and the chemicals that had more severe potential environmental and health impacts (benzene and BTEX). We also aimed to minimize the use of green as it is often associated with positive environmental messages and to assign similar colors to related chemicals.

This process suggests the importance of future research on cost-effective adaptable hardware-based color solutions for data physicalization and methods for color selection for limited palettes. The constraints of our project forced us to prioritize our need for cost-effective, accessible, and adaptable lighting that could work at night which limited our ability to take advantage of the extensive color research. Fortunately, our aims focused on bringing the community together around the data display to create a space for discussion and action and were not centered on in-depth visual interpretation of the data, which may minimize the need for more complex color representations.

*Experience design:* Our situated data physicalization on the Chelsea Creek aimed to provide ambient context for the data, better embody the information, and to create space for community conversation [15, 69, 87]. Though the data display is at a considerable distance from each individual facility, we consider it to be situated as it allows people to contextualize the water quality permit violations in the community through viewing the facilities as well as the river, homes, businesses, and people they impact. This context provides two benefits of situated displays described earlier:

"allow[ing] viewers to examine and extract additional information not present in the dataset itself" and using the physical environment's "potential to act as an information-carrying medium" [12, 87]. This location also met GreenRoots' goals of increasing community ownership of the waterfront, which is greatly limited because of the area's status as a Designated Port Area reserved for water-based industry [49].

GreenRoots was able to get access to a nearby dock that was safe for a large group of people and full of visual cues about the area. Building community reflection into the event was crucial to the action aims of PAR and is detailed in 4.3. We also consulted with the Boston University Cyberlaw Clinic, the Coast Guard, and the Chelsea police to ensure that they were aware of our event and that it met legal and safety standards. Specifically, we were concerned about the risk of copyright or libel laws associated with using the company logos on the lanterns in public and any limitations or permitting requirements for public events occurring on the water at night in a Designated Port Area. This was of particular concern as it was unclear how this event would be perceived by industries who have significant power locally and nationally.

### 4.3 Deployment

Roughly 60 people attended the Chemicals in the Creek event and/or the community meeting, which made it one of the larger community meetings held by GreenRoots, and the Facebook event video had over 250 views as of April 2020. During the event, each ECO student led the release of the lanterns for one of the seven oil storage facilities with help from a volunteer who they instructed.

Community participants gathered at GreenRoots before the event and were guided to the location of the data physicalization where they were given an event brochure (Fig. 2a). Shortly after sunset, the event was introduced by an ECO student and an academic researcher with interpretation to Spanish from a bilingual MIT undergraduate, as part of GreenRoots' language justice practices [79] (Fig. 2b). After a group countdown to begin the event, volunteers lit the 2013 sign (Fig. 2c) and put the lanterns representing the violations from that year onto the river (Fig. 2d). After 20 seconds, the next year was announced, the sign lit up, and the lanterns from that year added to the water. This process continued through 2017 as participants took in the experience from the dock (Fig. 2e). Once all the lanterns were in the water, community collaborators spontaneously spoke to the group about the violations and their meaning in the context of local environmental justice efforts. The community was invited to interact with the lanterns and volunteers began to remove them from the water (Fig. 2f). Participants returned to GreenRoots for a collective conversation which included questions about the data, brainstorming around potential actions, and post-it note feedback (Fig. 1d).

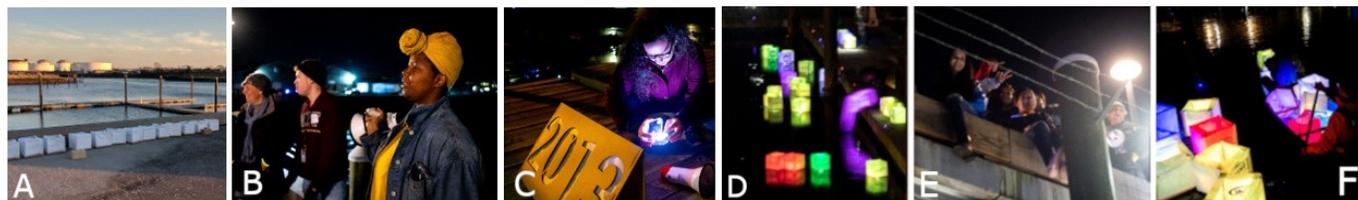

Fig. 2. After the (a) event set-up (img: Laura Perovich), Chemicals in the Creek began with (b) an introduction by a member of ECO, (img: Rio Asch Phoenix) until (c) the first year sign was lit up (img: Rio Asch Phoenix), and (d) the lanterns representing the violations from that year were put in the river (img: Will Campbell) as (e) community members observed (img: Rio Asch Phoenix). GreenRoots spoke to the crowd and (f) volunteers and event observers pulled the lanterns from the water before the community discussion (img: Rio Asch Phoenix).

## 5 FINDINGS: OUTCOMES

**Community experience:** *People felt that the information was accessible, visible, engaging, and in touch with the community.* They thought that "the lights were very good way to show the contamination" (C33) and felt that they "really learned new info about what's going on in the Creek" (C54). Participants also noted that the information was well explained and connected to the community, saying "I like how in touch it was with Chelsea residents and explaining the waste in the creek clearly" (C57) and that the event was "very inclusive to the community because they were first explaining 'we're going to give Spanish translation as well as English' and that was very helpful to help connect the whole community" (D10). Others remarked on the deep engagement during the community meeting discussion about the physicalization, saying "I was standing up in the front more, so I saw people's faces. They were really paying attention and absorbing this" (V05). This was echoed in observations from researchers and community partners and event photos (Figs. S1.08, S1.11, S1.14, S2.50-51).

At the same time, many people said that warmer or better weather would improve their experience (C24, C60, C35, V02, V09, V07, D13). The researchers also noted that the cold weather took away from the experience and made conversations outside before the event somewhat rushed. These physical aspects of designing experiences around situated data displays are important to surface in developing design practices for the space developed by Willett et al. [87].

**Types of understanding & affect:** *Participants described the main messages of the data physicalization and expressed positive affect about learning this through the event. Some provided more detailed comments indicating that they understood the data encoding.* Participants felt that the data physicalization expanded their knowledge, remarking that it was "interesting that each lantern represented different chemicals too, because it kind of showed, like it's not just one source of pollution" (V02). At times this information was eye opening for them: "we looked at like, which companies were responsible for what, but like, I didn't know any of that. So like seeing it, I was like 'oh my gosh, like, that's kind of crazy,'...that was really good for me" (V04). These large scale interpretations of the data reflected GreenRoots' priority of conveying big picture messages centered on community relevance in relation to collective action. This stands in contrast to many traditional analysis tools that aim to create extremely detailed understanding of data. This finding also supports previous results in Information Visualization around the power of artistic and physical modes of display in creating both emotional and intellectual engagement [25, 33, 36, 77, 86].

Six event volunteers or observers (V01, V03, V08, V02, V06, C26) used information about the data encoding (each lantern is a violation, each color is a chemical, the logo is the company, the time of release is the year) in interviews or comments. This provided preliminary evidence that some people understood the fundamental display structure. Our PAR methods and evaluation approach did not allow us to assess the breadth and depth of participant understanding.

**Physical interaction with situated data displays:** *The situated nature and physicality of Chemicals in the Creek supported the project aims and physical interaction with the lanterns was particularly salient.* People remarked that the location of the installation helped in communicating the information, "the contaminants were polluting the water so it kind of makes sense to have the water and show [how] the violations just creep up" (E14). Others noted that the presence of residential houses near the water provided important context for the data and that it was "good being there" (V09). These observations support the benefits of situated data outlined in [87] and the project goals of making open government data present and actionable in the community. The physicality of the event and the dynamism it provided also created a collective excitement, "It was fluid. It wasn't static. It wasn't just on the screen. It was data visualization that you could interact with and other people were excited about it as well. It was pretty dope" (V07).

The embodied nature of the event was particularly impactful for event helpers who "had to get down on our knees and reach down" to put the lanterns in the water which was "a moving experience... thinking about what this symbolized and what this meant and the full weight of what has been put into the water there. It was meaningful" (V01). At the same time, some event observers wanted a more physically engaged experience of the installation. An ECO member noted that of their friends who came to the event "the only problem one person had was, it wasn't as hands on for them. They thought they will be able to hold the lanterns to release them" (E18), a point echoed by another ECO member (E17). This points to the challenge of balancing the planning needs and structures of a ceremony with the desire of participants to spontaneously engage through embodied participation. We attempted to incorporate physical engagement with the lanterns by encouraging event observers to interact with the lanterns as they floated on the Creek, but it seems that did not meet their needs. Putting the lanterns in the water was a powerful moment and the event may have been improved by restructuring it to allow more people to participate in this.

**Performance:** *The aesthetics and performative nature of the event made it memorable and the tensions between the meaning of the data and the appearance of the display were thought provoking.* Many community members indicated that they liked how colorful the installation was (C34, C47, C56, C26, C49). Others enjoyed the "dramatic" (C27), "creative" (C25), and "unique" (C55) nature of the physicalization and found it to be a "memorable visual of pollution" (C28). Participants liked the light-up brochures (V08, D10) noting that they were "beyond amazing" (V09). The visual performative nature of the event drew a local newspaper that ran a full page photo article [60] and it was later covered by other local news outlets [24] and in a newsletter for Environmental Health researchers [54].

One participant noted that the attractiveness of the performance helped people look at information that they might have otherwise turned away from, "The contrast [between the aesthetics and the message] was really nice. I think that that really matters because a lot of people in the community, they might think it doesn't matter, they just don't look at it, but here in this event, you were forced to, because it was so bright" (V07). Others discussed how this tension between the visual experience of the display and the discomfort of the violations they represented provided an important point for conversation, "it was a very interesting contrast between this very really aesthetically pleasing visual experience representing this subject matter of toxins, and things that are detrimental to health. That contrast made talking about it a little easier I think to some people" (D13). This was echoed by another participant who noted how this tension sparked social interaction, "a lot of us talked about...the fact that the whole show was like, gorgeous, like looking at all the lights and it was really cool, but like the idea of it is like, really disappointing the fact that like, that's happening." (V04).

The use of performance in public spaces for engagement and conversation builds on findings on making visualizations public [11, 69], performance in HCI [75], and work on community building towards activism using art and technology in city spaces [22, 47]. At the same time, the people at the installation were largely community members, not environmental scientists, regulators, or industry workers. It is possible that a performative event would be received differently by these groups or would not fit their needs. For example, people from these groups may want to spend extended amounts of time with data and to easily sort, rank, or manipulate it, which was not possible in the fleeting nature of the performance.

**Designing a social experience around the data:** *Supporting conversations and the logistics of engagement with the data display are crucial in creating a positive experience and making the information seem actionable and accessible.* A number of participants noted that the community conversation after the event was central to reinforcing the collective experience of the situated data physicalization and making it actionable, "I liked the event and then how afterwards we all went back and then spoke about it...it also solidified more of the purpose, and why we're doing this, which was nice" (V07). Another participant noted that the conversations about the data physicalization at the community meeting were "very productive and very meaningful" (D15). Many previous urban installations have observed that public data physicalizations prompted social interaction [11, 12, 89]. This work builds from that foundation by creating more intentional spaces for interaction supported by community groups working on local areas of interest.

At the same time, four participants (C23, C36, C43, C61) noted that the event would have benefitted from being "more organized." As organizers, we noticed that the logistics of getting community members from GreenRoots to the dock could have been better crafted, though the installation itself was carefully choreographed. ECO suggested that we could spend more time talking with the community before showing the data physicalization, "I guess looking back we probably should have taken a little more time maybe…to tell the people who showed up what the problem is, but maybe a little more clearly because I think it felt a little rushed" (E14).

**PAR, participation, and project implementation:** *The ways in which community members and researchers participated varied throughout the project and it was not always clear what type of involvement was most appropriate at each phase*. The "ideation" and "deployment" phases of the design process were done very collaboratively by researchers and community partners, but community partners and researchers participated in different ways in the "implementation" phase of the project. In some cases, this was due to structural or logistical limitations. For example, the initial testing of the floating objects occurred in winter and therefore had to take place at an indoor pool which was only available for research when ECO was in school. Some fabrication tools were also only available at universities to trained students, which impacted the steps of the building process that ECO could contribute to. The researchers got feedback from ECO through regularly shared video and pictures but this engagement differs from a hands-on experience.

Other portions of the physical fabrication process were very time consuming and not central to the interests of ECO students or the core action aims of the project. This included extensive design testing to ensure that the lanterns would float and were cost effective. The researchers worked with a high school student from Chelsea that regularly spoke with ECO to stay connected in this phase. Some ECO students also found the lantern building process difficult and tedious, saying "we definitely need more help on [making the lanterns] 'cause after two hours of doing all that I'm really tired of this" (E17). Expanding involvement in the building of the lanterns could have the additional benefit of involving more people in the embodied experience of physically creating the data display. Such a large scale collective fabrication of community data displays could build from findings on individual fabrication of personal data which have facilitated deep reflection on both data and values [33, 77].

Balancing the "research" aspect of PAR with the "participatory" and "action" components led to different types of participation in implementing the data display. Researchers were more concerned with formalizing the data physicalization design (for example, detailing the design metaphor and the task and data abstraction) and the community was very active in developing the vision, aims, and outcomes of the project based on their data communication priorities. The graduation of the ECO students after the ideation process also created a divide between the project ideation and implementation. The presence of a continuing student and a series of introductory workshops helped bridge that gap, but participation throughout the entire process might have given a more complete experience.

These variations in roles is a known attribute of PAR processes. While "collaboration at all stages of reflection and action is ideal, it is important to recognise that levels of participation by co-researchers and participants may vary significantly… participants… may not desire full participation and care needs to be taken to work with people on their own terms" [40, 41]. At the same time, active hands-on collaboration can provide an important space for co-learning and discovery. Therefore, it may be difficult to know what level and type of interaction is most beneficial at each project phase.

**PAR, information design, and changing power relationships:** *Our PAR methodology contributed to building community ownership of the violation data and event outcomes, pride in the event, and began to engage with issues of power embedded in the NPDES data.* ECO students felt ownership of the event—they stated that they would be interested in doing a similar event again and felt they had the skills necessary to do so (E20). They were proud of the outcomes they achieved, eager to share them with the community, and pleased at the engagement of their peers, "I don't think we were expecting as much youth to show, 'cause sometimes when we host events they just come see the posts and don't go. But this time a youth group from East Boston, they had a lot of youth come, like seven or ten, and they were really interested and asking questions" (E20). Other members of ECO were glad their friends came (E12), saying "One thing I was really happy for was when the teens came by, and I'm like, oh my God, they actually came by to listen, they actually want to know about this." (E18) (Fig. S2.49). ECO was able to use this experience for further activism, including leading a public presentation [48] and organizing an Environmental Justice Youth Summit [20]. In these events, they highlight the power of their experience "being the ones to teach the adults rather than adults teaching youth" and say "we are the future, so we speak up now."

The engagement that ECO observed from the local youth was evident in the broader community during the meeting as participants raised powerful new questions. Community members explored the systems behind the violations, asking "What role does Massachusetts Department of Environmental Protection have in water protection?" (C31) which begins to surface the structures of power and paths to action on these issues. They also attempted to locally contextualize the data "How many violations have occurred in the same period on the island end (Mystic River) side of Chelsea?" (C32). Others also suggested that the event could be strengthened by "[getting] an elected official to observe [the event] and send [a] press release on your behalf" (C45) or otherwise engaging politicians (C59), indicating their desire to hold accountable those in power.

The community had control over the event outcomes which was important as the project aimed to increase community access to open government data and their ability to leverage it for change. After the event, GreenRoots discussed the possible causes of the violations with oil storage facility representatives at two meetings with support from researchers (Figs. S2.56-57). GreenRoots was able to use the event to create new pathways of communication for receiving this data. Representatives from the oil storage facilities informally agreed to increase direct communication with GreenRoots about NPDES violations, and more than one oil storage facility has followed through with this by emailing GreenRoots information about violations directly. This creates a new, more effective path for GreenRoots to access data that is important to their community and to gain information not found in the NPDES system, such as industry follow-up around the cause of the violation.

## 6 DISCUSSION

**Situated data displays, representational accuracy, and data witnessing as ceremony:** We have become accustomed to representing physical phenomena with numbers as part of the process of evaluating the significance of physical events [71]. Yet there is abstraction at work in considering these representation of pollution as "accurate" [62, 71]. Numbers may not surface the embodied experience of the violation: the waste flowing from a pipe into a public waterway and impacts on the health of the Chelsea Creek. Representing these violations through glowing lanterns in the community and on the river can remind us of the physicality of pollution and show the people and environments it negatively impacts. This can create space for collective reflection on possible actions and facilitate community ownership of information through hands on engagement. This builds from prior work showing that physical and contextual modes of representing data are accessible to a broader audience and provide space for emotional and social reflection [11, 12, 15, 16, 25, 36, 86, 87].

The lanterns and lights recall contemplative moments of togetherness, such as community vigils, that encourage deep collective attention over a defined but fleeting period of time. This approach distinguishes them from integrated urban data displays that invite but do not demand witnessing [11, 12, 69]. The use of lanterns to represent pollution data builds from work on the power of charismatic data "to spur specific managerial actions can address the newly visible problem" [62]. Turning CWA violations into a ceremonial event in the place where they occurred offers a shared space for recollection, evaluation, and expression of care for the Chelsea Creek and the lives it enriches [44]. What if every industry and community had to experience such ceremonies periodically? Would we think differently about the consequences of consumption systems that depend upon the production of toxic wastes [91]?

**PAR, modes of impact, and design methodology:** We see preliminary evidence that PAR approaches in data visualization design studies can lead to impacts that are owned by the community, and fit community needs [53]. In Chemicals in the Creek this was manifest in the community's ability to leverage the impact of the tool to change the paradigm. Specifically, GreenRoots has been able to use this data and the installation to advocate for themselves and gain more immediate and interpretable access to the data. At the same time, it is important to recognize that change in complex systems is difficult to achieve and this work is only a piece of the process.

Our study serves as a real world application of data display design through PAR in response to the call for civic data visualization methods that are deeply engaged with communities [12]. It builds from the use of co-design and participatory design methods in civic information displays [69, 89], PAR in HCI [27], and evaluation methods for visualization that center affective and social engagement [86]. It also contributes to expanding design study methodology for data visualization research [52] by incorporating action and social relevance. This is significant as researchers have noted how *research* communities tend to be informed by a particular political and cultural perspective that may remain invisible to them in their work, yet is central to the outcomes they create [90]. Dourish says that within the field of environmental computer science, researchers "[transform] the problem of sustainability into the cost-benefit trade offs of rational actor economics, promoting sustainability as a matter of personal morality rather than industrial regulation or political mobilization" [18]. Our study informed by PAR breaks this pattern by approaching environmental issues through the lens of regulation and political action, an approach that could be expanded in the information visualization community.

At the same time, PAR approaches can be time and resource intensive. Similarly, creating data physicalizations can be more resource intensive than creating data visualization [36]. A project that uses both these methods may face significant challenges. We experienced this in the cost of the resources for Chemicals in the Creek and the considerable logistical efforts needed for the event, for example the difficulties of storing and moving 76 lanterns. Holding the data physicalization outside offered additional challenges including changing tidal conditions, limited work possibilities in the winter months, and cold weather conditions during the event.

**Opportunities in data feminism research:** Feminist theory informs PAR [51]. Recently, researchers have proposed the idea of data feminism: "a way of thinking about data, both their uses and their limits, that is informed by direct experience, by a commitment to action, and by intersectional feminist thought" [15, 16]. This builds on feminist HCI [2, 3] and critical Information Visualization [17], among others [21, 64, 84]. Two of data feminism seven principles—"examine power" and "challenge power"—are central to PAR research, and two others—"elevate emotion and embodiment" and "consider context"—dovetail with research on data physicalization and new ways to evaluate data displays [15, 36, 86, 87]. Our work points towards the potential for data physicalization researchers to engage with data feminism and explore social issues increasingly of interest in Information Visualization.

**Limitations and future work:** This study focuses on one community and one open government dataset. PAR methods often lead to resource and time intensive collaboration which can preclude broad engagement and generalization of findings [85]. Fortunately, this methodology offers a greater depth of engagement that can lead to deeper project impact and adoption. To facilitate future studies, we have developed and shared tutorial videos for our process [60].

Our modes of evaluation focused on the overall aim of community action and emotional, social, and physical modes of evaluation of data visualizations [86]. Because of this, we did not explore the breadth and depth of understanding that participants gained from the data display or questions around perception. Because the event was designed for and advertised in the Chelsea community, many participants had prior relationships with GreenRoots or interest in environmental topics that may have led to a positive bias in their experience. Environmental scientists, regulators, and industry officials were not included in this study but may have different needs or priorities which may not be met by data performances.

## 7 CONCLUSION

We present a two-year design study of Chemicals in the Creek, a situated data physicalization that explores ways to engage communities with "open" government data drawing on performative and PAR methods. This work provides a real world application of situated data physicalization [87], builds on work that points towards the need for deeply engaged community-based design study methods that can address social issues [12, 52, 69], and centers the importance of designing the full experience of the data physicalization. Our PAR framework suggests a research process that centers community impacts, works towards re-embodying and re-contextualizing data, and facilitates community ownership. In doing so, it provides an application within the data visualization community of methods that may be able to create different types of impacts [52, 70]. Future work will explore how physical participation can be extended to more people while maintaining the coherence of the data performance and how situated data performances are interpreted by other groups such as field experts.

## ACKNOWLEDGEMENTS

Thanks to ECO and Leilani Mroczkowski who were central to this work. Thanks to event participants, the Chelsea community, Paddle Boston, OBM, and the BU Cyber Law Clinic. Thanks to contributing students: Michael Still, Gustavo Santiago-Reyes, Jacqueline Chen, Maggie Zhang, Xavier Mojica, Emily Schachtele, Garance Malivel, and Shawn Sullivan; event volunteers: John Rao, Arushi Sood, Ed Hackett, Sharon Harlan, Olivia Ozkurt, Abbie Keane, Holly Coppes, Hanson Au, Lourdes Vera, Marc Jacobson, Dorian Stump, Angela Stewart, Laura Senier, Grace Poudrier, and Kaline Langley; and photographers: Rio Asch Phoenix, Will Campbell, Jimmy Day, and David Mussina. Thanks to the IEEE InfoViz reviewers. Research support from CRESSH, Media Lab Elements, RIELS, and Harvard TH Chan School of Public Health JPB fellowship.

## REFERENCES

[1] P. Aoki, A. Woodruff, B. Yellapragada, and W. Willett, "Environmental protection and agency: Motivations, capacity, and goals in participatory sensing," in *Proceedings of the 2017 CHI Conference on Human Factors in Computing Systems*, 2017, pp. 3138–3150.

[2] S. Bardzell, "Feminist HCI: taking stock and outlining an agenda for design," in *Proceedings of the SIGCHI conference on human factors in computing systems*, 2010, pp. 1301–1310.

[3] S. Bardzell and J. Bardzell, "Towards a feminist HCI methodology: social science, feminism, and HCI," in *Proceedings of the SIGCHI Conference on Human Factors in Computing Systems*, 2011, pp. 675–684.

[4] R. Bhargava and C. D'Ignazio, "Designing tools and activities for data literacy learners," in *Workshop on Data Literacy, Webscience*, 2015.

[5] J. Bird and Y. Rogers, "The pulse of tidy street: Measuring and publicly displaying domestic electricity consumption," in *Pervasive 2010*, 2010.

[6] T. Blascheck, L. M. Vermeulen, J. Vermeulen, C. Perin, W. Willett, T. Ertl, and S. Carpendale, "Exploration Strategies for Discovery of Interactivity in Visualizations," *IEEE transactions on visualization and computer graphics*, vol. 25, no. 2, pp. 1407–1420, 2019.

[7] M. Boychuk, M. Cousins, A. Lloyd, and C. MacKeigan, "Do We need Data Literacy? Public Perceptions Regarding Canada's Open Data Initiative," *Dalhousie Journal of Interdisciplinary Management*, vol. 12, no. 1, 2016.

[8] M. Brehmer, S. Ingram, J. Stray, and T. Munzner, "Overview: The design, adoption, and analysis of a visual document mining tool for investigative journalists," *IEEE transactions on visualization and computer graphics*, vol. 20, no. 12, pp. 2271–2280, 2014.

[9] J. M. Chevalier and D. J. Buckles, *Participatory action research: Theory and methods for engaged inquiry*. Routledge, 2019.

[10] S. Claes, J. Coenen, and A. V. Moere, "Conveying a civic issue through data via spatially distributed public visualization and polling displays," in *Proceedings of the 10th Nordic Conference on Human-Computer Interaction*, 2018, pp. 597–608.

[11] S. Claes and A. V. Moere, "The role of tangible interaction in exploring information on public visualization displays," in *Proceedings of the 4th International Symposium on Pervasive Displays*, 2015, pp. 201–207.

[12] S. Claes and A. Vande Moere, "Street infographics: raising awareness of local issues through a situated urban visualization," in *Proceedings of the 2nd ACM International Symposium on Pervasive Displays*, 2013, pp. 133–138.

[13] A. Crabtree, A. Chamberlain, R. E. Grinter, M. Jones, T. Rodden, and Y. Rogers, *Introduction to the special issue of "The Turn to The Wild."* ACM New York, NY, USA, 2013.

[14] T. Davies and M. Frank, "'There's no such thing as raw data': exploring the socio-technical life of a government dataset," in *Proceedings of the 5th Annual ACM Web Science Conference*, 2013, pp. 75–78.

[15] C. D'Ignazio and L. F. Klein, *Data Feminism*. The MIT Press, 2020.

[16] C. D'Ignazio and L. F. Klein, "Feminist data visualization," in *Workshop on Visualization for the Digital Humanities (VIS4DH), Baltimore. IEEE*, 2016.

[17] M. Dörk, P. Feng, C. Collins, and S. Carpendale, "Critical InfoVis: exploring the politics of visualization," in *CHI'13 Extended Abstracts on Human Factors in Computing Systems*, 2013, pp. 2189–2198.

[18] P. Dourish, "Print this paper, kill a tree: Environmental sustainability as a research topic for human-computer interaction," *Submitted to Proc CHI*, 2009.

[19] P. Dutta, P. M. Aoki, N. Kumar, A. Mainwaring, C. Myers, W. Willett, and A. Woodruff, "Common sense: participatory urban sensing using a network of handheld air quality monitors," in *Proceedings of the 7th ACM conference on embedded networked sensor systems*, 2009, pp. 349–350.

[20] EDGI Comms, "EJxYouth Summit Speakers," *Environmental Data and Governance Initiative*, 29-Nov-2019. [Online]. Available: envirodatagov.org/ejxyouth-summit-speakers-we-arent-waiting-for-adults-were-leading-the-way/. [Accessed: 31-Jul-2020].

[21] S. Elwood, "Volunteered geographic information: future research directions motivated by critical, participatory, and feminist GIS," *GeoJournal*, vol. 72, no. 3–4, pp. 173–183, 2008.

[22] J. Fredericks, "From Smart City to Smart Engagement: Exploring Digital and Physical Interactions for Playful City-Making," in *Making Smart Cities More Playable*, Springer, 2020, pp. 107–128.

[23] P. Freire and D. Macedo, *Pedagogy of the Oppressed*, 30th Anniversary edition. New York: Continuum, 2000.

[24] A. Gaffin, "Lanterns on Chelsea Creek highlight ongoing chemical problems," *Universal Hub*, 22-Apr-2019. [Online]. Available: https://www.universalhub.com/2019/lanterns-chelsea-creek. [Accessed: 21-Apr-2020].

[25] P. Gough, C. de Berigny Wall, and T. Bednarz, "Affective and effective visualisation: Communicating science to non-expert users," in *2014 IEEE Pacific Visualization Symposium*, 2014, pp. 335–339.

[26] M. B. Gurstein, "Open data: Empowering the empowered or effective data use for everyone?," *First Monday*, vol. 16, no. 2, 2011.

[27] G. R. Hayes, "The relationship of action research to human-computer interaction," *ACM Transactions on Computer-Human Interaction (TOCHI)*, vol. 18, no. 3, p. 15, 2011.

[28] J. Hendler, J. Holm, C. Musialek, and G. Thomas, "US government linked open data: semantic. data. gov," *IEEE Intelligent Systems*, vol. 27, no. 3, pp. 25–31, 2012.

[29] L. Hespanhol and M. Tomitsch, "Power to the People: Hacking the City with Plug-In Interfaces for Community Engagement," in *The Hackable City*, Springer, Singapore, 2019, pp. 25–50.

[30] U. Hinrichs, S. Carpendale, N. Valkanova, K. Kuikkaniemi, G. Jacucci, and A. V. Moere, "Interactive public displays," *IEEE Computer Graphics and Applications*, vol. 33, no. 2, pp. 25–27, 2013.

[31] C. Huisingh, "Burden of Asthma in Massachusetts," Massachusetts Department of Public Health, 2009.

[32] S. Huron, S. Carpendale, A. Thudt, A. Tang, and M. Mauerer, "Constructive visualization," in *Proceedings of the 2014 conference on Designing interactive systems*, 2014, pp. 433–442.

[33] S. Huron, P. Gourlet, U. Hinrichs, T. Hogan, and Y. Jansen, "Let's Get Physical: Promoting Data Physicalization in Workshop Formats," in *Proceedings of the 2017 Conference on Designing Interactive Systems*, 2017, pp. 1409–1422.

[34] R. Jacobs, S. Benford, M. Selby, M. Golembewski, D. Price, and G. Giannachi, "A conversation between trees: what data feels like in the forest," in *Proceedings of the SIGCHI Conference on Human Factors in Computing Systems*, 2013, pp. 129–138.

[35] Y. Jansen and P. Dragicevic, "An interaction model for visualizations beyond the desktop," *IEEE Transactions on Visualization and Computer Graphics*, vol. 19, no. 12, pp. 2396–2405, 2013.

[36] Y. Jansen, P. Dragicevic, P. Isenberg, J. Alexander, A. Karnik, J. Kildal, S. Subramanian, and K. Hornbaek, "Opportunities and challenges for data physicalization," in *Proceedings of the 33rd Annual ACM Conference on Human Factors in Computing Systems*, 2015, pp. 3227–3236.

[37] S. Kestin and J. Maines, "Speeding Cops: A Sun Sentinel Investigation," 2012. [Online]. Available: https://www.sun-sentinel.com/news/speeding-cops/. [Accessed: 25-Mar-2019].

[38] S. Kim, E. Paulos, and M. D. Gross, "WearAir: expressive t-shirts for air quality sensing," in *Proceedings of the fourth international conference on Tangible, embedded, and embodied interaction*, 2010, pp. 295–296.

[39] S. Kim, C. Robson, T. Zimmerman, J. Pierce, and E. M. Haber, "Creek watch: pairing usefulness and usability for successful citizen science," in *Proceedings of the SIGCHI Conference on Human Factors in Computing Systems*, 2011, pp. 2125–2134.

[40] S. Kindon, R. Pain, and M. Kesby, *Participatory action research approaches and methods: Connecting people, participation and place*, vol. 22. Routledge, 2007.

[41] R. Kitchin, "Using Participatory Action. Research Approaches in Geographical Studies of Disability: Some Reflections," *Disability Studies Quarterly*, vol. 21, no. 4, pp. 61–69, 2001.

[42] L. Koeman, V. Kalinkaitė, Y. Rogers, and J. Bird, "What chalk and tape can tell us: lessons learnt for next generation urban displays," in *Proceedings of the international symposium on pervasive displays*, 2014, p. 130.

[43] S. Kuznetsov, G. N. Davis, E. Paulos, M. D. Gross, and J. C. Cheung, "Red balloon, green balloon, sensors in the sky," in *Proceedings of the 13th international conference on Ubiquitous computing*, 2011, pp. 237–246.

[44] M. P. de La Bellacasa, *Matters of care: Speculative ethics in more than human worlds*, vol. 41. U of Minnesota Press, 2017.

[45] G. Lane, C. Brueton, G. Roussos, N. Jeremijenko, G. Papamarkos, D. Diall, D. Airantzis, and K. Martin, "Public Authoring & Feral Robotics," *Proboscis. Cultural Snapshot Number Eleven*, 2006.

[46] V. Lim, E. Deahl, L. Rubel, and S. Williams, "Local Lotto: Mathematics and mobile technology to study the lottery," in *Cases on technology integration in mathematics education*, IGI Global, 2015, pp. 43–67.

[47] C. Liu, M. Balestrini, and G. N. Vilaza, "From social to civic: Public engagement with iot in places and communities," in *Social Internet of Things*, Springer, 2019, pp. 185–210.

[48] G. Malivel, "'Environmental Data Justice: Vision and Values' Event," *Environmental Data and Governance Initiative*, 24-May-2019. [Online]. Available: https://envirodatagov.org/environmental-data-justice-vision-and-values-event/. [Accessed: 31-Jul-2020].

[49] Massachusetts Office of Coastal Zone Management, "CZM Port and Harbor Planning Program," *Mass.gov*, 2020. [Online]. Available: https://www.mass.gov/service-details/czm-port-and-harbor-planning-program-designated-port-areas. [Accessed: 07-Nov-2018].

[50] J. McCarthy and P. Wright, "Technology as experience," *interactions*, vol. 11, no. 5, pp. 42–43, 2004.

[51] A. McIntyre, *Participatory Action Research*. Los Angeles: Sage, 2007.

[52] M. Meyer and J. Dykes, "Criteria for rigor in visualization design study," *IEEE transactions on visualization and computer graphics*, vol. 26, no. 1, pp. 87–97, 2019.

[53] M. Minkler and N. Wallerstein, Eds., *Community-Based Participatory Research for Health: From Process to Outcomes*, 2 edition. San Francisco, CA: Jossey-Bass, 2008.

[54] NIEHS, "PEPH Newsletter," *National Institute of Environmental Health Sciences*, Aug-2019.

[55] B. Nissen and J. Bowers, "Data-things: digital fabrication situated within participatory data translation activities," in *Proceedings of the 33rd Annual ACM Conference on Human Factors in Computing Systems*, 2015, pp. 2467–2476.

[56] D. Offenhuber, "Data by Proxy-Material Traces as Autographic Visualizations," *IEEE transactions on visualization and computer graphics*, 2019.

[57] D. Offenhuber, "The Invisible Display–Design Strategies for Ambient Media in the Urban Context," *International Workshop on Ambient Information Systems, Colocated with Ubicomp*, p. 152, 2008.

[58] A. Parker and S. Dosemagen, "Environmental protection belongs to the public: A vision for citizen science at EPA," in *AGU Fall Meeting Abstracts*, 2017.

[59] L. J. Perovich, S. Wylie, and R. Bongiovanni, "Pokémon Go, pH, and projectors: applying transformation design and participatory action research to an environmental justice collaboration in Chelsea, MA," *Cogent Arts & Humanities*, p. 1483874, 2018.

[60] L. Perovich, S. A. Wylie, and R. Bongiovanni, "Open Water Data," *Open Water Project*, 2019. [Online]. Available: http://datalanterns.com/. [Accessed: 25-Mar-2019].

[61] L. Perovich, S. A. Wylie, and R. Bongiovanni, "GitHub: data lanterns," 07-Feb-2019. [Online]. Available: github.com/lperovich/dataLanterns. [Accessed: 26-Mar-2019].

[62] K. H. Pine and M. Liboiron, "The politics of measurement and action," in *Proceedings of the 33rd Annual ACM Conference on Human Factors in Computing Systems*, 2015, pp. 3147–3156.

[63] Pittsburgh Art Places and A. Polli, "Particle Falls," 2008. [Online]. Available: http://www.pittsburghartplaces.org/accounts/view/1000. [Accessed: 20-May-2020].

[64] M. Posner, "What's Next: The Radical, Unrealized Potential of Digital Humanities," *Miriam Posner's Blog*, 27-Jul-2015. [Online]. Available: http://miriamposner.com/blog/whats-next-the-radical-unrealized-potential-of-digital-humanities/. [Accessed: 21-Apr-2020].

[65] President Obama, *Executive order–making open and machine readable the new default for government information*. 2013.

[66] President Obama, *Memorandum on Transparency and Open Government*. 2009.

[67] Publications Office of the European Union, "European Union Open Data Portal," 2020. [Online]. Available: data.europa.eu/euodp/en/data/. [Accessed: 25-Mar-2019].

[68] F. Samsel, L. Deck, and B. Campbell, "Climate Prisms: The Arctic Connecting Climate Research and Climate Modeling via the Language of Art," *IEEE VIS Arts Program (VISAP)*, 2015.

[69] J. Schoffelen, S. Claes, L. Huybrechts, S. Martens, A. Chua, and A. V. Moere, "Visualising things. Perspectives on how to make things public through visualisation," *CoDesign*, vol. 11, no. 3–4, pp. 179–192, 2015.

[70] M. Sedlmair, M. Meyer, and T. Munzner, "Design study methodology: Reflections from the trenches and the stacks," *IEEE transactions on visualization and computer graphics*, vol. 18, no. 12, pp. 2431–2440, 2012.

[71] N. Shapiro, N. Zakariya, and J. Roberts, "A wary alliance: From enumerating the environment to inviting apprehension," *Engaging Science, Technology, and Society*, vol. 3, pp. 575–602, 2017.

[72] B. Shneiderman, "The eyes have it: A task by data type taxonomy for information visualizations," in *The Craft of Information Visualization*, Elsevier, 2003, pp. 364–371.

[73] R. Sosa, V. Gerrard, A. Esparza, R. Torres, and R. Napper, "Data Objects: Design Principles for Data Physicalisation," in *DS92: Proceedings of the DESIGN 2018 15th International Design Conference*, 2018, pp. 1685–1696.

[74] A. S. Taylor, S. Lindley, T. Regan, D. Sweeney, V. Vlachokyriakos, L. Grainger, and J. Lingel, "Data-in-place: Thinking through the relations between data and community," in *Proceedings of the 33rd Annual ACM Conference on Human Factors in Computing Systems*, 2015, pp. 2863–2872.

[75] R. Taylor, J. Spence, B. Walker, B. Nissen, and P. Wright, "Performing research: Four contributions to HCI," in *Proceedings of the 2017 CHI Conference on Human Factors in Computing Systems*, 2017, pp. 4825–4837.

[76] A. Thudt, U. Hinrichs, and S. Carpendale, "Data craft: integrating data into daily practices and shared reflections," *CHI 2017 Workshop on Quantified Data & Social Relationships*, 2017.

[77] A. Thudt, U. Hinrichs, S. Huron, and S. Carpendale, "Self-reflection and personal physicalization construction," in *Proceedings of the 2018 CHI Conference on Human Factors in Computing Systems*, 2018, p. 154.

[78] US EPA, "Enforcement and Compliance History Online," 2019. [Online]. Available: https://echo.epa.gov/. [Accessed: 25-Mar-2019].

[79] US EPA, "Environmental Justice Analysis in Support of the National Pollutant Discharge Elimination System (NPDES) Permits for the Chelsea River Bulk Petroleum Storage Facilities.," Region 1, 2014.

[80] US EPA, "Air Quality Data Collected at Outdoor Monitors Across the US," 08-Jul-2014. [Online]. Available: www.epa.gov/outdoor-air-quality-data. [Accessed: 25-Mar-2019].

[81] US Government, "Data.gov," *Data.gov*, 2019. [Online]. Available: https://www.data.gov/. [Accessed: 25-Mar-2019].

[82] N. Valkanova, S. Jorda, and A. V. Moere, "Public visualization displays of citizen data: design, impact and implications," *International Journal of Human-Computer Studies*, vol. 81, pp. 4–16, 2015.

[83] F. B. Viegas, M. Wattenberg, F. Van Ham, J. Kriss, and M. McKeon, "Manyeyes: a site for visualization at internet scale," *IEEE transactions on visualization and computer graphics*, vol. 13, no. 6, pp. 1121–1128, 2007.

[84] J. Wajcman, "Feminist theories of technology," *Cambridge journal of economics*, vol. 34, no. 1, pp. 143–152, 2010.

[85] N. Wallerstein, B. Duran, M. Minkler, and J. G. Oetzel, *Community-based participatory research for health: advancing social and health equity*. John Wiley & Sons, 2017.

[86] Y. Wang, A. Segal, R. Klatzky, D. F. Keefe, P. Isenberg, J. Hurtienne, E. Hornecker, T. Dwyer, and S. Barrass, "An emotional response to the value of visualization," *IEEE computer graphics and applications*, vol. 39, no. 5, pp. 8–17, 2019.

[87] W. Willett, Y. Jansen, and P. Dragicevic, "Embedded data representations," *IEEE transactions on visualization and computer graphics*, vol. 23, no. 1, pp. 461–470, 2017.

[88] A. Wolff, D. Gooch, J. Cavero, U. Rashid, and G. Kortuem, "Removing barriers for citizen participation to urban innovation," in *The Hackable City*, Springer, 2019, pp. 153–168.

[89] N. Wouters, J. Huyghe, and A. Vande Moere, "OpenWindow: citizen-controlled content on public displays," in *Proceedings of the 2nd ACM International Symposium on Pervasive Displays*, 2013, pp. 121–126.

[90] S. A. Wylie, *Fractivism: Corporate bodies and chemical bonds*. Duke University Press, 2018.

[91] S. Wylie, N. Shapiro, and M. Liboiron, "Making and Doing Politics Through Grassroots Scientific Research on the Energy and Petrochemical Industries," *Engaging Science, Technology, and Society*, vol. 3, pp. 393–425, 2017.